\documentclass{jpsj-suppl}
\usepackage{txfonts} 

\title{Double Resonance in Dalitz Plot of $M_{p\Lambda}$-$M_{K\Lambda}$ in DISTO Data on $p+p\rightarrow p+\Lambda+K^+$ at 2.85 GeV}

\author{K. Suzuki$^{1}$, T. Yamazaki$^{2,3}$, M. Maggiora$^{4}$ and P. Kienle$^{1,5}$ for the DISTO collaboration}

\inst{$^{1}$Stefan Meyer Institute for Subatomic Physics, Austrian Academy of Sciences, Boltzmanngasse 3, 1090, Vienna, AUSTRIA,\\
$^{2}$Department of Physics, University of Tokyo, Tokyo, 116-0033 Japan,\\
$^{3}$RIKEN Nishina Center, Wako, Saitama, 351-0198 Japan,\\
$^{4}$Dipartimento di Fisica Generale "A. Avogadro" and INFN, Torino, Italy,\\
$^{5}$Excellence Cluster Universe, Technische Universit\"at M\"unchen, Garching, Germany}

\email{ken.suzuki@oeaw.ac.at}

\recdate{Feb 23 ty, 2015, talk given at Int Conf Hyp2015, Sendai, in press in JPSJ-suppl}

\abst{The $X(2265)$ resonance was previously observed in DISTO data of $p+p\rightarrow p+\Lambda+K^+$ at 2.85 GeV on an attempt of searching for the kaonic nuclear state $K^-pp \rightarrow p + \Lambda$. In the present paper we report an additional finding, namely, a double resonance type phenomena, not only with a peak at $M(p\Lambda) = 2265$ MeV/$c^2$ but also a broad bump at $M(K^+ \Lambda) \sim 1700$ MeV/$c^2$. This "double-resonance" zone is expressed as $XY(2265, 1700)$. The latter bump may result from nearby nucleon resonances, typically $N^*(1710)$, as well as by attractive $K^+ - \Lambda$ final-state interaction. We point out that this double resonance $XY(2265, 1700)$ as seen in DISTO at 2.85 GeV cannot be populated kinematically in a HADES experiment at 3.5 GeV.} 

\kword{kaonic nuclear state, strange dibaryon, $K^-pp$, $X(2265)$, double resonance $XY(2265, 1700)$  }

\begin{document}
\maketitle

\section{Introduction}

The DISTO collaboration has reported~\cite{Yamazaki:10, Maggiora:10} in  the $p+p\rightarrow p+K^++\Lambda$ at 2.85 GeV reaction an observation of a statistically significant broad resonance with baryon number 2, strangeness $-1$ with its mass centered around 2267 MeV/c$^2$ (aka $X(2265)$), ~100 MeV below the threshold.
The analysis was performed on an attempt of searching for a kaonic nuclear bound state, i.e. the most basic dibaryonic $\bar{K}NN_{S=0, I=1/2}=K^-pp$ system~\cite{Akaishi:02,Yamazaki:02}. This search followed the prediction \cite{Yamazaki:07} that it can be produced abundantly in $p + p \rightarrow K^+ + \Lambda^* + p \rightarrow K^+ + K^-pp$ reaction through the $\Lambda^* + p$ as a doorway, if the $K^-pp$ is exceptionally dense, as predicted, so as to be formed by extraordinary strong sticking of  $\Lambda^*$ and $p$ produced in the short-range $p-p$ collision. Here, $\Lambda^* \equiv \Lambda(1405) \equiv I=0~ K^-p$ plays an essential role. In fact, a successive $p-p$ experimental data taken at 2.5 GeV incident energy \cite{Kienle:12}, which is too low to produce $\Lambda^*$, demonstrated that the yield for the peak $X(2265)$ is diminished. 
@
The quest for the kaonic nuclear bound states has been pursued since the original prediction~\cite{Akaishi:02, Yamazaki:02} for over a decade, but except for DISTO \cite{Yamazaki:10, Maggiora:10}, which owed a special favorable reaction mechanism \cite{Yamazaki:07}, it was difficult to find some robust signature in singles spectra over quasi-free hyperon backgrounds. From "unsuccessful" observation one tends to believe that our object, $K^-pp$, {\it does not exist}, thus being confused between {\it its existence versus its  production}. Only recently a new tide of excitements with developed experimental configurations has become available. 

The E27 experiment at J-PARC reported a "$K^-pp$-like structure", which was observed to an extremely small level of signal to background ratio ($\sim 0.01$) in the $d(\pi^+, K^+)$ reaction at 1.69 GeV/c~\cite{Ichikawa:15}.
The $K^-pp$ peak was not expected to appear in singles spectra of the $d(\pi^+,K^+)$ reaction (see Fig.11 of \cite{Yamazaki:07}), and thus, they configured coincidence counter arrays, which helped to reveal a significant peak with its mass and width similar to the $X(2265)$.
@
On the other hand, the initial claim of an observation of a strange dibaryon by FINUDA in the stopped $K^-$ reaction on several light nuclear targets decaying to back-to-back $\Lambda$-$p$~\cite{Agnello:05} has not been reproduced, as reported by their updated data and analysis~\cite{Agnello:13,Filippi:14}. Some negative results of $K^-pp$ search have also been reported, using photoproduction~\cite{Tokiyasu:14}. One may say that the production of $\Lambda(1405)$ in the same reaction plays a key leading role as a doorway for $K^-pp$ formation. No sizable appearance of $\Lambda(1405)$ is seen in such "unsuccessful" experiments. 

Only one exception to this understanding might be the case of the $pp$ reaction at 3.5 GeV by HADES collaboration~\cite{Agakishiev:15}. We will come back to this point after discussing the main subject of the present paper.
Here, we discuss one new feature of the $X(2265)$ seen in the 2.85 GeV data, $M(p\Lambda)-M(K^+ \Lambda)$ {\it double resonance}, which could explain the non-observation of the $X(2265)$ in the HADES data~\cite{Agakishiev:15} at 3.5 GeV. We also find a follow-up paper~\cite{EF:15} by some of the authors of ~\cite{Agakishiev:15}, who hold doubts and criticisms on our analyses~\cite{Yamazaki:10, Maggiora:10, Kienle:12}. Those our replies to this paper besides the interpretation of their 3.5 GeV data will be published elsewhere~\cite{Suzuki:16}.

\section{Dalitz plot presentation of the $pp\rightarrow p\Lambda K^+$ reaction}

Figure 1 shows the Dalitz plot of the $pp\rightarrow p\Lambda K^+$ reaction at $T_p = 2.85$ GeV in $M(p\Lambda)$ vs $M(K^+\Lambda)$ presentation and their projections. Here, the mass-linear expression is taken for easier readability of the mass.
As described in Ref.~\cite{Yamazaki:10}, the acceptance uncorrected data was normalized bin by bin by the Monte Carlo data of pure phase space $p\Lambda K^+$ final states analyzed in an exactly same manner as the real data. In this way we expect a flat $M^2(p\Lambda)$ and $M^2(K^+\Lambda)$ distribution if the reaction is purely for phase space and on the other hand, if the reaction contains any resonances or final state interactions, the distribution will {\it deviate} from the flat distribution.
A proton angular cut ($|cos\theta_{cm}(p)|<0.6$) is applied to the data in order to suppress the ordinary $p\Lambda K^+$ decay process and to enhance relatively the $pp\rightarrow K^+ + X(2265)$ component~\cite{Yamazaki:10}.

The $M(p\Lambda)$ projection shows a distinct peak of $X(2265)$. In the $M(K^+\Lambda)$ distribution one sees also   a non-flat, structure peaked at around 1700 MeV/c$^2$ that may be attributed to a presence of $N^*$ resonances. Though several $N^*$ resonances could contribute to it, $N^*$(1710), which could be dominant at this energy~\cite{El-Samad:10}, is known to have a broad width of 50 to 250 ($\sim$ 100) MeV and to decay to $K^+$ and $\Lambda$ with a 5-25\% branching ratio~\cite{PDG}.

\begin{figure}[tbh]
\center
\includegraphics{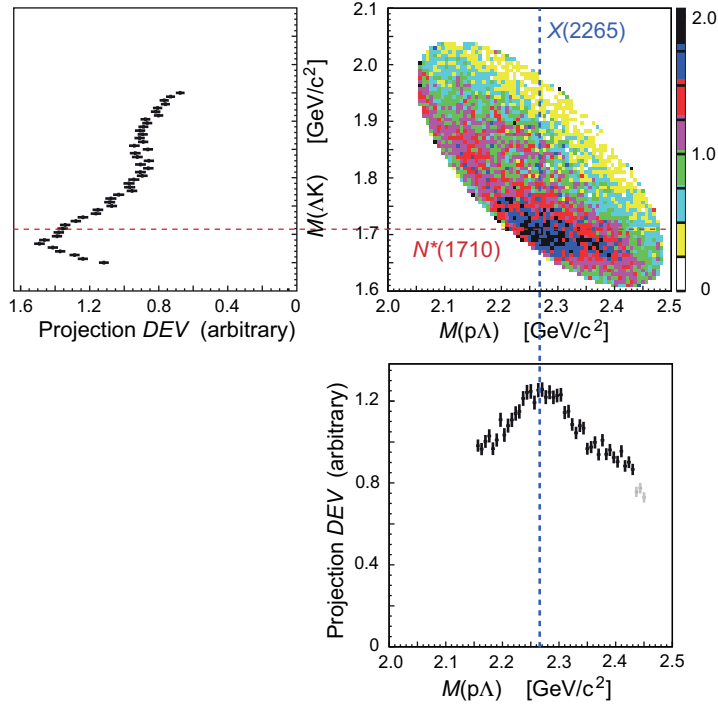}
\caption{Dalitz plot presentation of the $p+p\rightarrow X(2265)+K^+$ reaction at $T_p = 2.85$ GeV collected by DISTO collaboration in $M(p\Lambda)$ vs $M(K^+\Lambda)$~\cite{Yamazaki:10} and its projections. The data are presented as {\it deviation spectra} from the pure phase-space scenario, which is supposed to be flat (here, however, the flatness is approximate as the mass linear expression is taken).
Both projections show clear structures, $M(p\Lambda)\sim 2265$ GeV/c$^2$ and $M(K^+\Lambda)\sim 1700$ GeV/c$^2$.}
\label{f1}
\end{figure}

We consider the following two-step process with a population of $\Lambda(1405)$ to form the $X(2265)$, and a final state interaction between $\Lambda$ and $K^+$.

\begin{equation}
p+p\rightarrow (p+\Lambda^*)+K^+ \rightarrow X(2265)+K^+ \rightarrow p+\Lambda+K^+
\label{e1}
\end{equation}

The population of $N^*$ (or equivalently, attractive $\Lambda-K^+$ final state interaction) seems to favor the production of $X(2265)$ at the crossing point of two resonances. We may call this double resonance peak as $XY(2265, 1700)$.
Assuming Eq.~\ref{e1}, the sticking probability of $\Lambda(1405)$ and proton to form $X(2265)$ is found to be very high ($>0.5$). Possibly the resonance is pronounced by a joint effect of two resonances.
A similar double resonance phenomenon has been reported at low energy heavy-ion reactions~\cite{Shimoura:86}. In this case: $^{12}$C + $^{12}$C $\rightarrow$ $^{12}$C + $\alpha$ + $^{8}$Be, the intensity is peaked at the crossing point of two resonances, namely $^{8}$Be + $\alpha \leftrightarrow ^{16}$O and $\alpha + ^{12}$C $\leftrightarrow ^{12}$C.

\begin{figure}[tbh]
\center
\includegraphics[width=12cm]{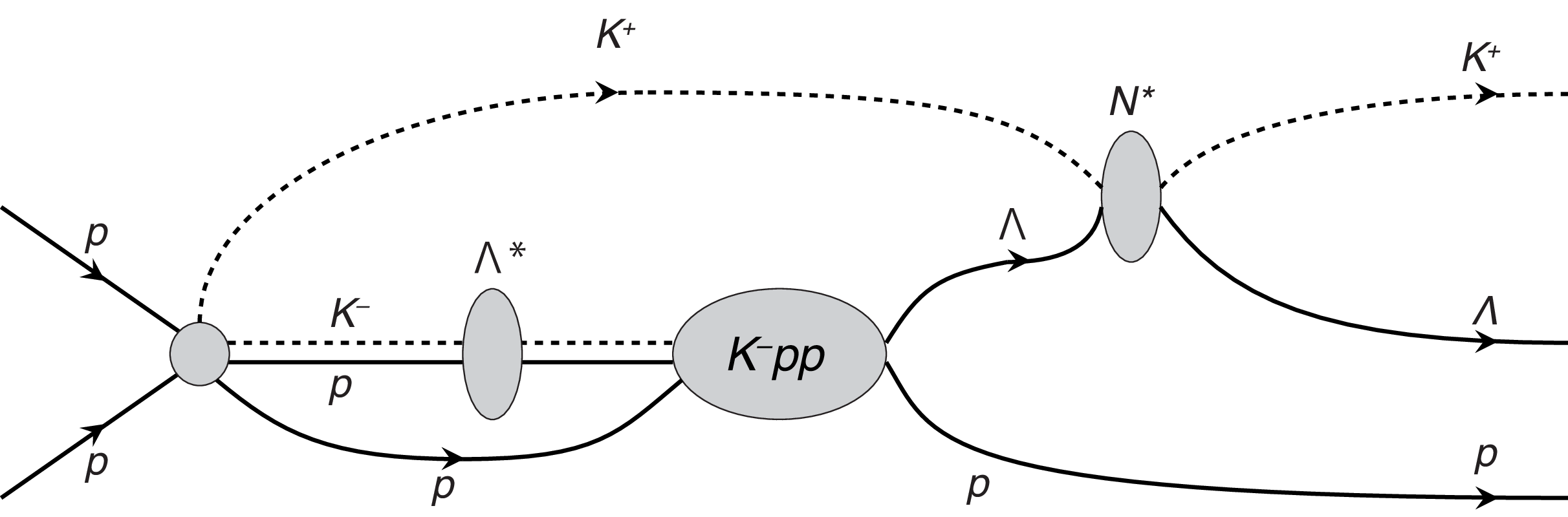}
\caption{Feynman diagram of the $pp\rightarrow p\Lambda K^+$ reaction with $X(2265)$ and $N^*$ double resonance.}
\label{f2}
\end{figure}

Figure~\ref{f2} depicts a possible Feynman diagram of such a reaction.

\section{Non-population of $XY(2265, 1700)$ at 3.5 GeV HADES collision}

\begin{figure}[tbh]
\center
\includegraphics[width=7cm]{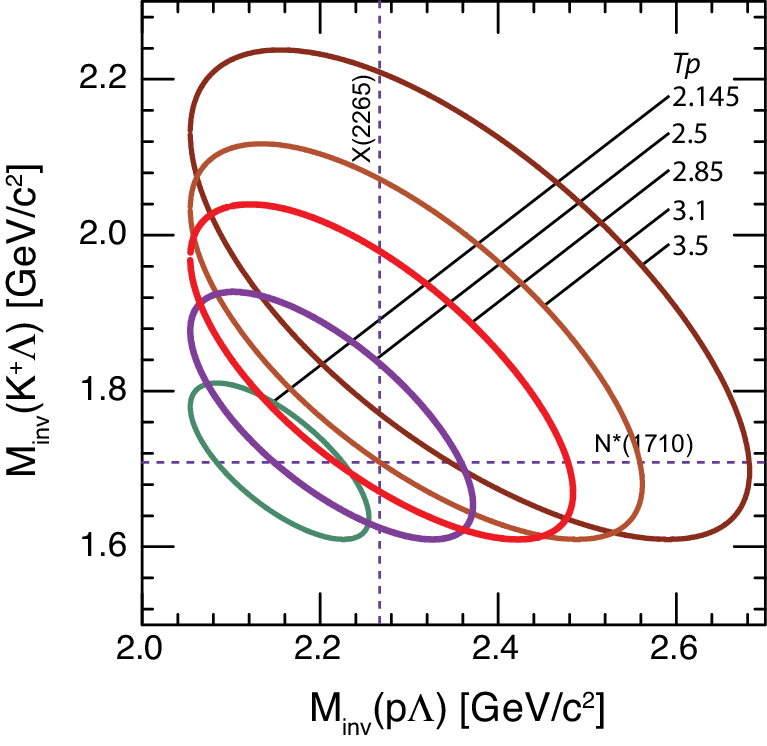}
\caption{Kinematically allowed Dalitz domain of the $pp\rightarrow p\Lambda K^+$ reaction at various incident proton kinetic energies. The two resonances observed in Ref.~\cite{Yamazaki:10} are indicated by horizontal and vertical dashed lines. The double resonance point sits outside of the Dalitz domain for the $T_p=3.5$ GeV case where the non observation of $X(2265)$ was reported.}
\label{f3}
\end{figure}

This leads to another interesting consequence which explains a non-observation of the $X(2265)$ resonance in the same $pp\rightarrow p\Lambda K^+$ reaction at a higher energy at 3.5 GeV by HADES~\cite{Agakishiev:15} and even higher energies~\cite{BNL, LRL}.
As seen in Fig.~\ref{f3}, the double resonance point of $X(2265)$ and $N^*$ is outside of the kinematically allowed oval Dalitz domain of $T_p=3.5$ GeV and higher.

We reported earlier~\cite{Kienle:12} that the $T_p=2.5$ GeV is too low for the production of $X(2265)$, since the energy is only marginally above the production threshold of $\Lambda(1405)$ that would be fused with proton to form a $X(2265)$. Together the choice of the $T_p=2.85$ GeV turned out to be very unique.

\section{Summary}

In the exclusive data of $pp\rightarrow p\Lambda K^+$ reaction at $T_p=2.85$ GeV, two resonances are observed, namely, $M(p\Lambda)\sim 2265$ GeV/c$^2 = X(2265)$ and $M(K^+\Lambda)\sim 1700$ GeV/c$^2$. The events are pronounced at the crossing point of these two resonances, namely, at $XY(2265, 1700)$. 
We discussed a possibility of two resonances that lead to a high production yield of the $X(2265)$ resonance. We also found the uniqueness of the 2.85 GeV proton incident energy of this reaction for the production of the $X(2265)$.



\end{document}